\documentclass[twoside,leqno,
]{factaBw}
\usepackage{amsmath}
\usepackage{latexsym,amssymb,amsfonts,euscript}
\usepackage[section]{placeins}
\usepackage{graphicx}
\input epsf

 \factanis{ \,(2018),  -- }   
 \Received{ , 2018.}     

 \AMSMSC{04.50.Kd, 04.80.Cc, 04.25.Nx, 04.50.-h}

\title{CONSTRAINING SCALAR-TENSOR GRAVITY MODELS BY S2 STAR ORBITS AROUND THE GALACTIC CENTER
\footnote{The authors were supported by Ministry of Education, Science and Technological Development of the Republic of Serbia, through the project 176003 ''Gravitation and the Large Scale Structure of the Universe'', and by Istituto Nazionale di Fisica Nucleare, Sezione di Napoli, Italy, iniziative specifiche TEONGRAV and QGSKY. The authors also acknowledge the support by Bilateral cooperation between Serbia and Italy 451-03-01231/2015-09/1 ''Testing Extended Theories of Gravity at different astrophysical scales'' and of the COST Action CA15117 (CANTATA), supported by COST (European Cooperation in Science and Technology). This work was also partially supported by ICTP--SEENET-MTP project NT-03 ''Cosmology -- Classical and Quantum Challenges''.}}

\author{Vesna Borka Jovanovi\'{c}, Predrag Jovanovi\'{c}, Du\v{s}ko Borka, \\
Salvatore Capozziello, Stefania Gravina, and Anna D'Addio}

\begin{document}

\maketitle

\abstract
The aim of our investigation is to derive a particular theory among the class of scalar-tensor(ST) theories of gravity, and then to test it by studying kinematics and dynamics of S-stars around supermassive black hole (BH) at Galactic Center (GC). We also  discuss the Newtonian limit of this class of ST theories of gravity, as well as its parameters. We compare the observed orbit of S2 star with our simulated orbit which we obtained theoretically with the derived ST potential and constrained the parameters. Using the obtained best fit parameters we calculated orbital precession of S2 star in ST gravity,and found that it has the same direction as in General Relativity (GR), but causes much larger pericenter shift.
\endabstract

\section{Introduction}
\label{section1}
 \vglue-10pt
 \indent

Modified theories of gravity have been proposed as alternative approaches to Newtonian gravity in order to cure shortcomings of Newtonian gravity and GR, but without introducing dark matter and dark energy \cite{capo12,noji11}. Huge number of alternative gravity theories have been proposed (see e.g. review papers \cite{clif06,capo11a,capo11b} and the book \cite{clif12}). All these theories have to be also checked by astronomical observations taken on different astronomical scales, from the Solar System, binary pulsars, elliptical and spiral galaxies to the clusters of galaxies and cosmological scales \cite{capo02,capo03,carr04,iori10,capo11a,capo11b,soti10,leon11,capo14}.

Extended theories of gravity \cite{capo11a,capo11b} are alternative theories of gravity developed from the similar starting points investigated first by Einstein and Hilbert, but instead of Ricci curvature scalar $R$, one assumes a generic function $f$ of the Ricci scalar $R$. Using extended theories of gravity in our previous papers we tried to explain different astrophysical phenomena like: orbital precession of S2 star \cite{bork12,bork13,capo14,zakh14,bork16}, fundamental plane of elliptical galaxies \cite{bork16,bork16a}, the baryonic Tully-Fisher relation of gas-rich galaxies \cite{capo17}, and also to give the mass constraints for graviton \cite{zakh16,zakh18}.

S-stars are the bright stars which move around the centre of our Galaxy where the compact radio source Sgr A$^\ast$ is located (more about this can be found in references \cite{ghez00,scho02,ghez08,gill09a,gill09b,genz10,gill12,meye12,gill17,hees17,chu17}). The progress in monitoring bright stars near GC has been made \cite{gill17,hees17,chu17}, but still the current astrometric limit is not sufficient to unambiguously confirm deviations of S2 star orbit from Keplerian one. We expect that future observations of S-stars will be more precise with astrometric errors several times smaller than currently are.

Some models of extended gravity, and in particular generic models containing ST and higher-order curvature terms, are described in \cite{capo15}. In this study, we consider possible signatures for a ST theory within the Galactic Central Parsec, not tested at these scales yet. Using gravitational potential that we derived from the modified theories of gravity \cite{dadd18,grav18}, we compare simulated and observed orbits of S2 star. In this way we are able to investigate the orbital precession of S2 star, deviations from the Keplerian orbit, the stellar kinematics around supermassive BH at GC, as well as to constrain parameters of the derived potential.

This paper is organized as follows: in \S 2 we explain theoretical base of ST gravity, in \S 3 we describe our two body numerical simulations, \S 4 is devoted to the obtained results and their discussion, and finally in \S 5 we point out main results of our study.

\section{Scalar-Tensor theory of gravity and its parameters}
\label{section2}
 \vglue-10pt
 \indent

In ST theory of gravity both, the metric tensor $g_{μν}$ and a fundamental scalar field $\phi$, are involved. This theory of gravity contains two arbitrary functions of the scalar field: the coupling $F(\phi)$ and the interaction potential $V(\phi)$. $F(\phi)$ underlines a non-minimal coupling between the scalar field and the geometry, and $V(\phi)$ implies a self-interaction of the field. More about general scalar-tensor Lagrangian see in \cite{capo96}.

We take the most general action in four dimensions of a theory of gravity where a scalar field is non-minimally coupled to the geometry of the form \cite{capo93,capo94}:

\begin{equation}
S = S_{M}+\dfrac{1}{2\kappa^2}\int{d^{4}x}\sqrt{-g}[F(\phi)R+\dfrac{3}{
2\phi}g^{\mu\nu}\phi_{,\mu}\phi_{,\nu}-V(\phi)],
\end{equation}

\noindent and choose a specific form for $F(\phi) = \xi \phi^m$, $V(\phi) = \lambda \phi^n$, where $S_M$ is the matter action, $\xi$ is a coupling constant, $\lambda$ gives the self-interaction potential strength, $m$ and $n$ are arbitrary parameters. We take a rather general choice for arbitrary functions $F(\phi)$ and $V(\phi)$ which is in agreement with the existence of a Noether symmetry \cite{capo93,capo94,capo96}. Also, several ST physical theories (e.g. induced gravity) admit such a form for $F(\phi)$ and $V(\phi)$.

We investigated few different cases for $h_{00}=0.5\Phi$, where $\Phi$ is Newtonian-like potential.

In the case of sferical symmetry, and for a point distribution of matter, the linearized equations have the solutions as follows \cite{grav17,dadd17}. In case of $n\neq 0$ and $n\neq 2m$, solution is:
\begin{equation}
h_{00} \simeq \dfrac{\kappa^2}{4\pi\xi \varphi_0^m
}\dfrac{M}{r}-\dfrac{ \lambda}{2\xi}\varphi_0^{n-m}r^2-\dfrac{
\kappa^2 m^2M}{3(1-m^2 \varphi_0^{m-1}\xi)}\dfrac{e^{-pr}}{4\pi r}.
\end{equation}

In case of $n=2m$, solution is:
\begin{equation}
h_{00} \simeq \dfrac{\kappa^2M}{4\pi
r}[\dfrac{3-3m^2\varphi_0^{m-1}\xi-m^2\xi\varphi_0^m}{
3\xi\varphi_0^m(1-m^2\varphi_0^{m-1}\xi)}]-\dfrac{\lambda\varphi_0^m}
{2\xi}r^2.
\end{equation}

In case of $n=1$, solution is:
\begin{equation}
h_{00} \simeq \dfrac{\kappa^2M}{4\pi
r}[\dfrac{3-3m^2\varphi_0^{m-1}\xi-m^2\xi\varphi_0^m}{
3\xi\varphi_0^m(1-m^2\varphi_0^{m-1}\xi)}]-\dfrac{\lambda\varphi_0^{
1-m}}{2\xi}r^2.
\end{equation}

The ST gravitational potential in the weak field limit can be written in the following form \cite{dadd18,grav18}:

\begin{equation}
U_{ST}=\dfrac{\widetilde{G}}{\xi \varphi_0^m }\dfrac{M}{r}-\dfrac{ 
\lambda}{4\xi}\varphi_0^{n-m}r^2-\dfrac{ \widetilde{G} m^2M}{3(1-m^2 
\varphi_0^{m-1}\xi)}\dfrac{e^{-pr}}{ r},
\end{equation}

\noindent where $\varphi_0$ is positive real number (close to 1), $p$ is function of the ST gravity parameters $\xi, \lambda, m, n$:

\begin{equation}
p=\sqrt{\frac{\lambda
n\varphi_{0}^{n-1}(2m-\lambda{n})}{3(m^2\xi\varphi_{0}^{m-1}-1)}},
\end{equation}

\noindent and $\widetilde{G}$ is related with a gravitation constant $G_N$:

\begin{equation}
\label{ }
\tilde{G}=-\bigg[
\dfrac{3(1-m^2\varphi_0^{m-1}\xi)\xi\varphi_0^m}{3-\xi(3m^2\varphi_0^
{m-1}+m^2\varphi_0^m)}\bigg]G_N.
\end{equation}

\section{Simulated S2 star orbits in ST and Newtonian potential}
\label{section3}
 \vglue-10pt
 \indent

In order to constrain the parameters observationally, we simulated orbits of S2 star in the modified gravitational potential, and then we compared the results with the set of S2 star observations obtained by the New Technology Telescope (NTT) and Very Large Telescope (VLT).

As S2 star is one of the brightest among S-stars, with the short orbital period and the smallest uncertainties in determining the orbital parameters, it is a good candidate for this study. We draw orbits of S2 star in ST and Newtonian potential. For that purpose, we performed two-body simulations in the modified ST gravity potential: $U_{ST}(r) = C_1 \cdot \dfrac{1}{r} + C_2 \cdot r^2 + C_3 \cdot \dfrac{e^{-pr}}{r}$, with $C_1 = C_1(\xi, m)$, $C_2 = C_2(\xi, \lambda, m, n)$, $C_3 = C_3(\xi, m)$, $p = p(\xi, \lambda, m, n)$, and in Newtonian potential: $U_N(r) = - \dfrac{GM}{r}$.

The equations of motion in ST gravity are:

\begin{equation}\label{eqnmoto}
\dot{\overrightarrow{r}}=\overrightarrow{v}, \hspace{2 cm}
\mu\ddot{\overrightarrow{r}}=-\overrightarrow{\bigtriangledown}U_{ST}
(\overrightarrow{r}),
\end{equation}

\noindent where $\mu={M\cdot m_S}/({M+m_S})$ is the reduced mass in the two-body problem.

\begin{figure}[ht!]
\centering
\includegraphics[width=0.45\textwidth]{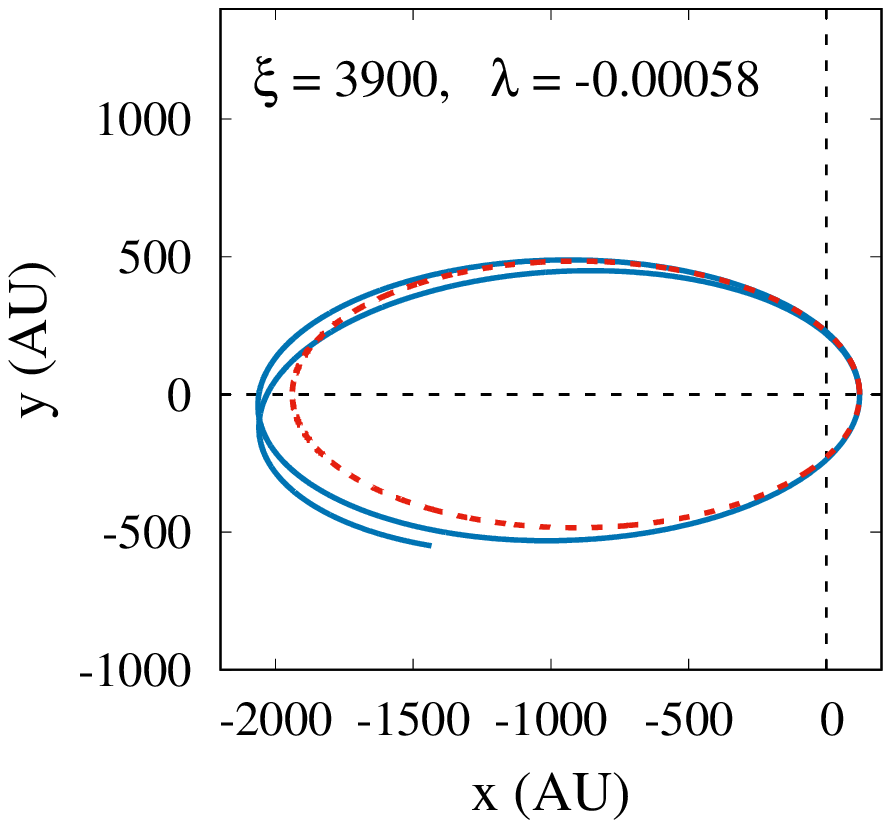}
\hspace{0.5cm}
\includegraphics[width=0.45\textwidth]{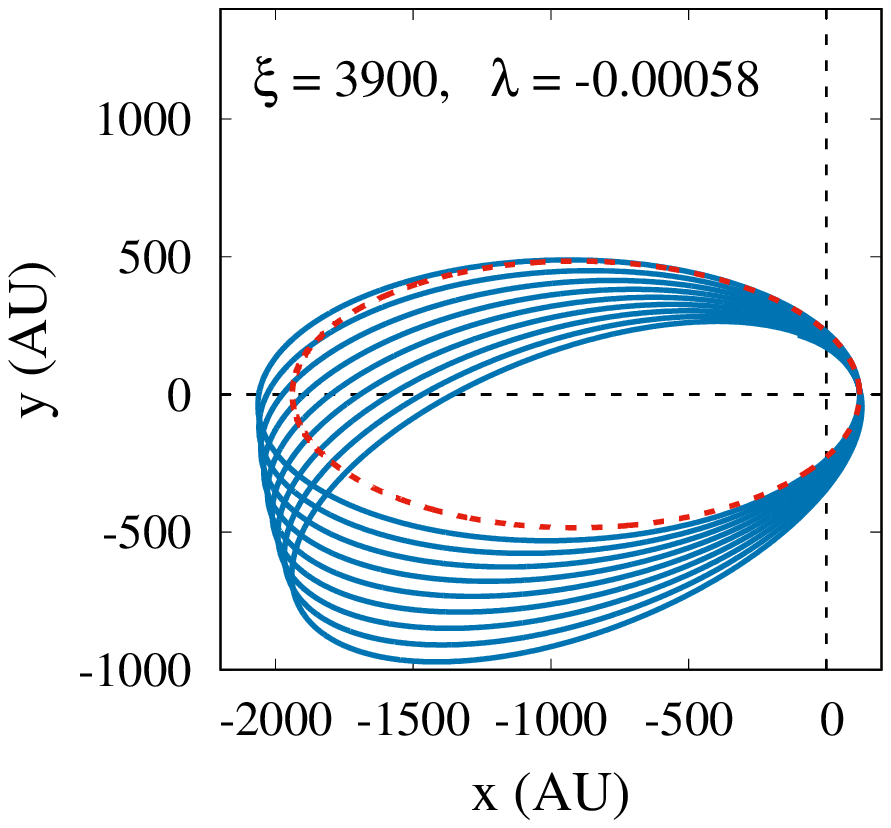}
\caption{Comparison between the orbit of S2 star in Newtonian potential (red dashed line) and ST potential (blue solid line) for parameters ($m$,$n$) = (1,3) and ($\xi$,$\lambda$) = (3900,-0.00058) during the time $t$ = 2T and 10T, where T is Keplerian period.}
\label{fig01}
\end{figure}

One example of the comparison between the orbit of S2 star in Newtonian and ST potential is given in Fig. \ref{fig01}. Our results show that there is a positive precession (as in GR), and after some number of periods the prograde shift results in rosette-shaped orbits.

We compare the obtained theoretical results for S2-like star orbits in the ST potential with the available set of observations of the S2 star. The observations, collected between 1992 and 2008 at the European Southern Observatory (ESO), with the optical telescopes NTT and VLT, are publicly available as the supplementary online data to the electronic version of Ref. \cite{gill09b}.

For comparing the astrometric observations with the fitted orbit, our method of calculation is the following:
\begin{itemize}
\setlength{\itemsep}{0em}
\item[$1^o$] we calculate the positions of S2 star in the orbital plane (the true orbit) by the numerical integration of equations of motion;
\item[$2^o$] then we project the true orbit to the observer's plane (the apparent orbit);
\item[$3^o$] we estimate the discrepancy between the simulated and the observed apparent orbit by the reduced $\chi^2$ \cite{bork13}.
\end{itemize}

\begin{figure}[ht!]
\centering
\includegraphics[width=0.60\textwidth]{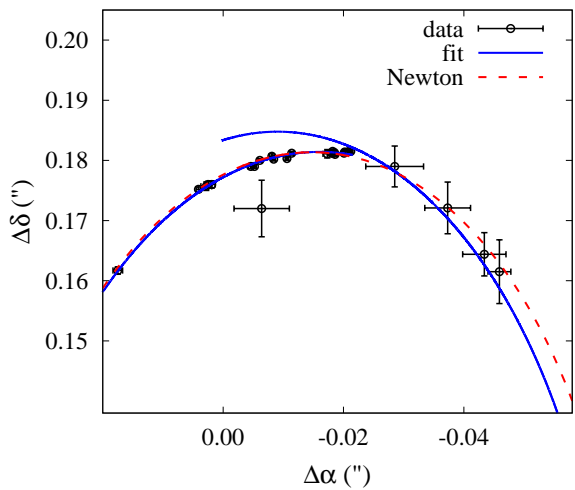}
\caption{Comparison of the NTT/VLT astrometric observations (black circles) and the fitted orbit in ST modified gravity (blue solid line) of S2 star around the Galactic Center, for ST gravity parameters ($m$,$n$) = (1,3) and ($\xi$,$\lambda$) = (3900,-0.00058). The Newtonian orbit is added with red dashed line.}
\label{fig02}
\end{figure}

From the comparison of the observations and the fitted orbit of S2 star around the GC, it can be clearly seen that the precession exists. In Fig. \ref{fig02} we present one part of the orbit near apocenter where orbital precession is obvious.

\section{Results and discussion}

\subsection{Constraints on ST gravity parameters}
\label{section4}
 \vglue-10pt
 \indent

\begin{figure}[ht!]
\centering
\includegraphics[width=0.95\textwidth]{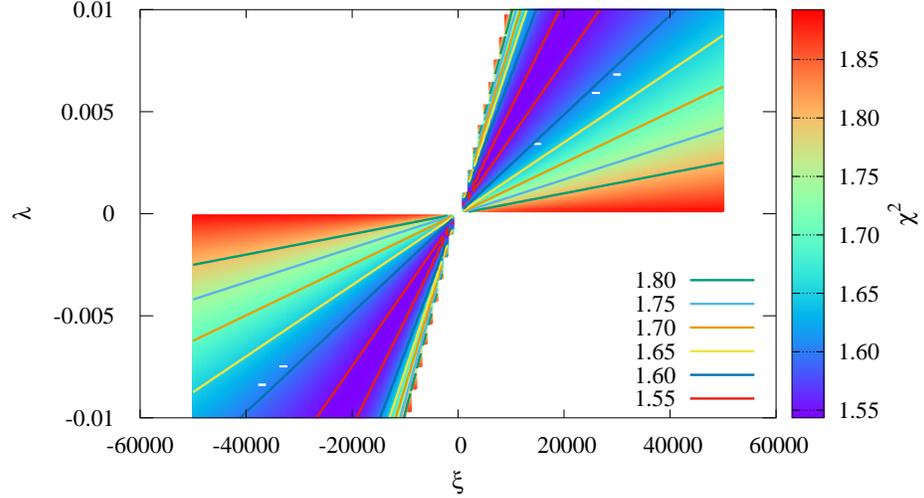}
\caption{The map of the reduced $\chi^2$ over the parameter space $(\xi, \lambda)$ of ST gravity in case of NTT/VLT observations of S2 star which give at least the same or better fits ($\chi^2 \le 1.89$) than the Keplerian orbits. Figure represents case for $(m,n)=(1,1)$. A few contours are presented for specific values of reduced $\chi^2$ given in the bottom right part of the figure.}
\label{fig03}
\end{figure}

\begin{figure}[ht!]
\centering
\includegraphics[width=0.95\textwidth]{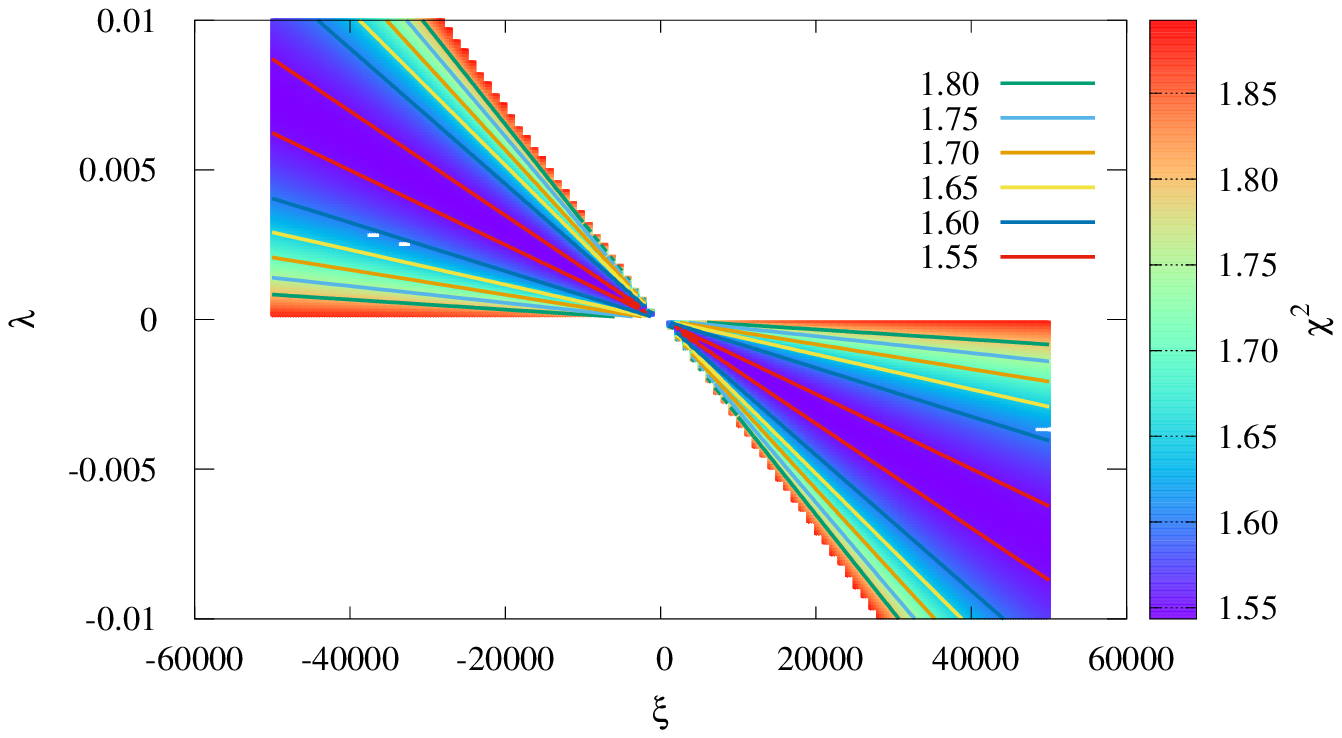}
\caption{The same as Fig. \ref{fig03}, but for case $(m,n)=(1,3)$.}
\label{fig04}
\end{figure}

\begin{figure}[ht!]
\centering
\includegraphics[width=0.95\textwidth]{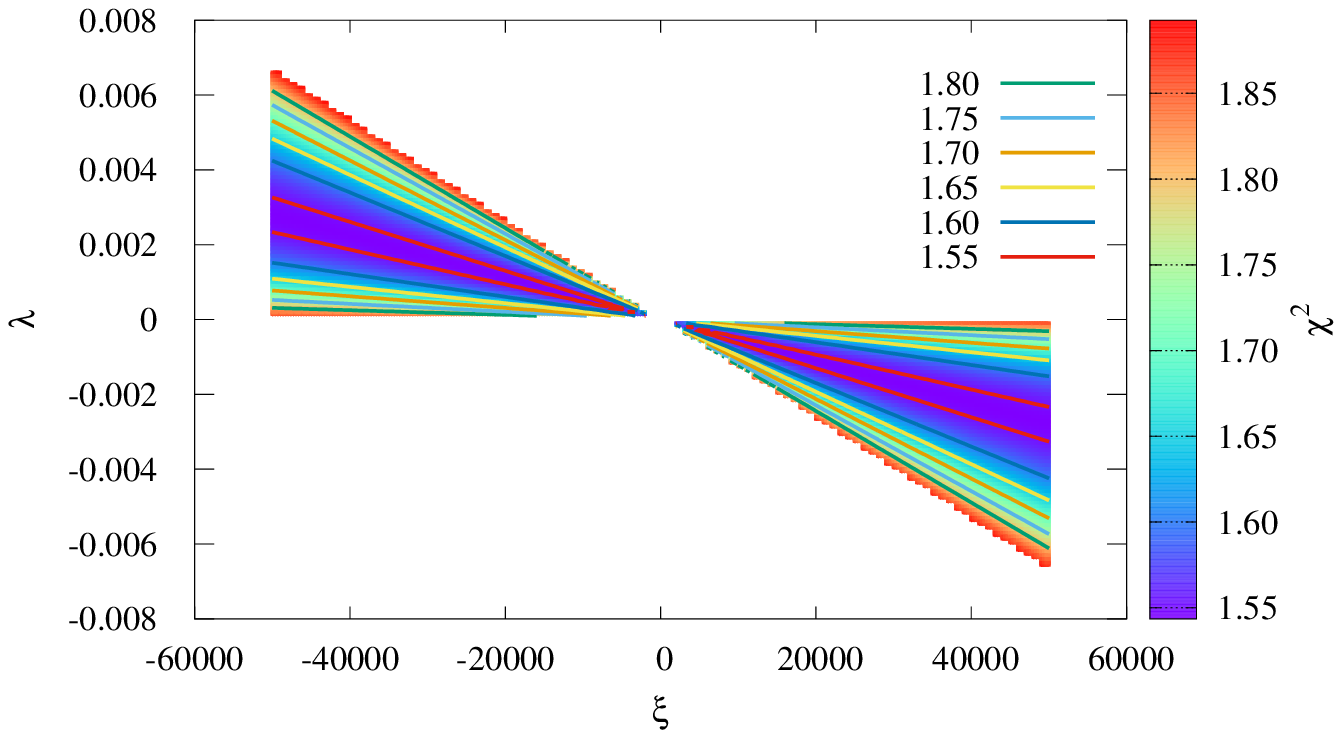}
\caption{The same as Fig. \ref{fig03}, but for case $(m,n)=(1,4)$.}
\label{fig05}
\end{figure}

\begin{figure}[ht!]
\centering
\includegraphics[width=0.95\textwidth]{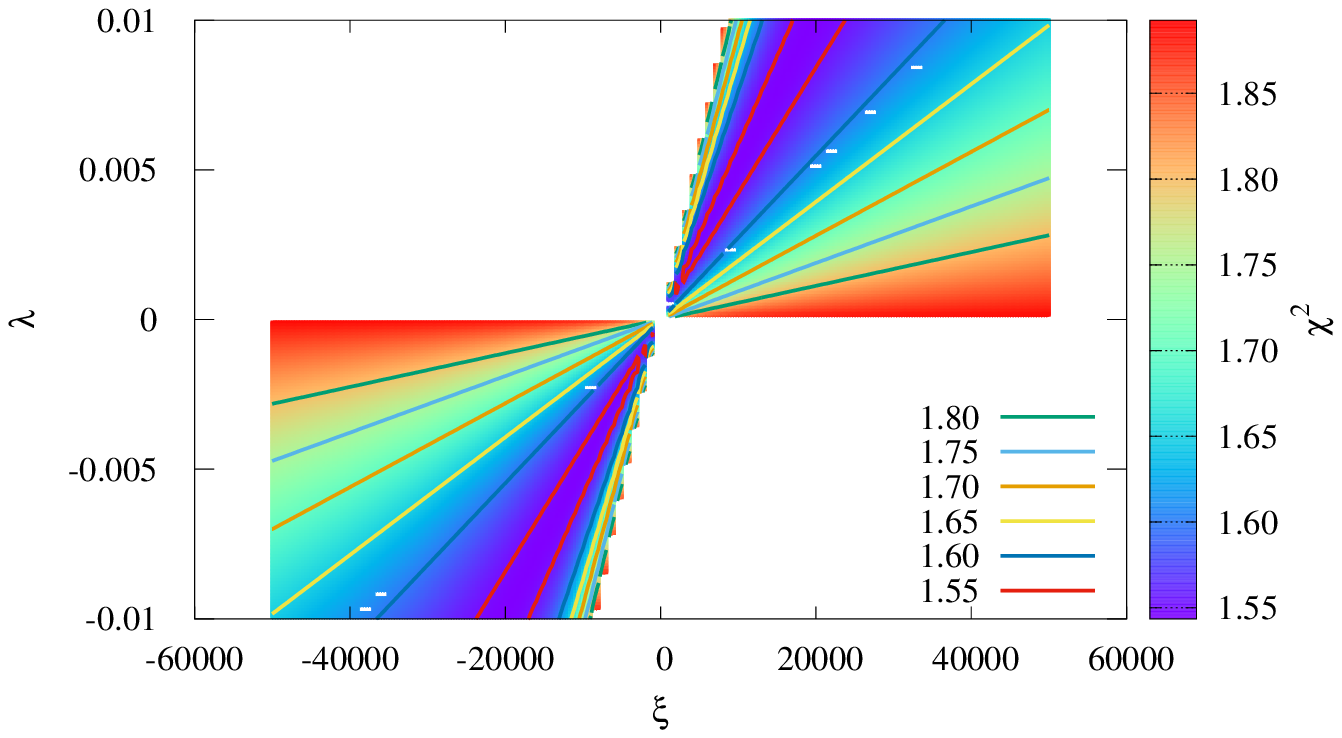}
\caption{The same as Fig. \ref{fig03}, but for case $(m,n)=(3,4)$.}
\label{fig06}
\end{figure}

For constraining ST gravity parameters, we choose some values for ($m$,$n$), and vary the parameters ($\xi$,$\lambda$) over some intervals, and search for those solutions which for the simulated orbits in ST gravity give at least the same or better fits ($\chi^2 \le 1.89$) than the Keplerian orbits. Then, we repeat the procedure for different combinations $(m,n) \rightarrow [1,10]$. Some maps of the reduced $\chi^2$ over the parameter space ($\xi$,$\lambda$) of ST gravity, for different combinations of $m$ and $n$, we show in Figs. \ref{fig03}-\ref{fig06}. The calculated $\chi^2_{min}$ values and the corresponding best fit values $\xi \underline{~}_{min}$ and $\lambda \underline{~}_{min}$ are given in Table \ref{tab01}. The readers should pay attention here that $\chi^2_{min}$ is the minimal value, but the parameters $\xi \underline{~}_{min}$, $\lambda \underline{~}_{min}$ are not minimal, but the best fit values which correspond to $\chi^2_{min}$.

As it can be seen from Figures \ref{fig03}-\ref{fig06}, as well as from Table \ref{tab01}, different combinations of $m$ and $n$ parameters give different best fit values for $\xi$ and $\lambda$, but they will not significantly affect the resulting orbital precession, as it will be shown below.

\begin{table}
\caption{Best fit values for ST gravity parameters, for different combinations of $m$ and $n$ (we take $\varphi_0$ = 1).}
\vglue4mm
\centering
{
\begin{tabular}{|c|c|c|r|r|}
\hline
$m$ & $n$ & $\chi^2_{min}$ & $\xi \underline{~}_{min}$ & $\lambda \underline{~}_{min}$ \\
\hline
1 & 1 & 1.5434350 & 13000 & 0.0058 \\
1 & 3 & 1.5434440 & 43000 & -0.0064 \\
1 & 4 & 1.5434426 & 43000 & -0.0024 \\
2 & 1 & 1.5434345 & 16000 & 0.0095 \\
2 & 2 & 1.5434352 & 15000 & 0.0067 \\
2 & 3 & 1.5434474 & -1000 & -0.0006 \\
3 & 1 & 1.5434336 & 1000 & 0.0008 \\ 
3 & 2 & 1.5434383 & 1000 & 0.0005 \\
3 & 3 & 1.5434352 & 15000 & 0.0067 \\
3 & 4 & 1.5434383 & 1000 & 0.0005 \\
4 & 1 & 1.5434317 & 4000 & 0.0041 \\
4 & 2 & 1.5434478 & -1000 & -0.0006 \\
4 & 3 & 1.5434348 & 21000 & 0.0100 \\
4 & 4 & 1.5434353 & 15000 & 0.0067 \\
10 & 10 & 1.5434353 & -15000 & -0.0067 \\ 
\hline
\end{tabular}}
\label{tab01}
\end{table}

\subsection{Orbital precession estimates in ST gravity}
\label{section5}
 \vglue-10pt
 \indent

In order to calculate orbital precession in ST modified gravity, under assumption that ST potential does not differ significantly from Newtonian potential, we derived perturbing potential:

\begin{equation}
 V(r)=U_{ST}-U_N; \hspace{2 cm} U_N=-\dfrac{GM}{r}.
\end{equation}

The obtained perturbing potential is of the form:

\begin{equation}
 V(r)= -\dfrac{GM}{r}\dfrac{\xi m^2
\varphi_0^m}{3-\xi(3m^2\varphi_0^{m-1}+m^2\varphi_0^m}-\dfrac{
\lambda}{4\xi}\varphi_0^{n-m}r^2-\dfrac{ \widetilde{G} m^2M}{3(1-m^2
\varphi_0^{m-1}\xi)}\dfrac{e^{-pr}}{ r}.
\end{equation}

\noindent and it can be used for calculating the precession angle according to Eq. (30) in Ref. \cite{adki07}:

\begin{equation}
\Delta \theta = \dfrac{-2L}{GM e^2}\int\limits_{-1}^1 {\dfrac{z \cdot
dz}{\sqrt{1 - z^2}}\dfrac{dV\left( z \right)}{dz}},
\label{prec}
\end{equation}

\noindent where $r$ is related to $z$ via: $r = \dfrac{L}{1 + ez}$. By differentiating the perturbing potential $V(z)$ and by substituting its derivative and $L = a\left( {1 - {e^2}} \right)$ into (\ref{prec}), we can obtain value for precession angle. Some calculated values are given in Table 1 from \cite{grav18}.

The precession of S2 star orbit is in the same direction with respect to GR and produces a prograde shift that results in rosette-shaped orbits. The pericenter advances by 2.5$^\circ$ per orbital revolution, while in GR the shift is 0.18$^\circ$.

\section{Conclusions}
\label{section6}
 \vglue-10pt
 \indent

As the S2 star is one of the brightest among S-stars, with the short orbital period and the smallest uncertainties in the orbital parameters, we find that it is a good candidate for this study. First we obtained the parameter space $(\xi,\lambda)$ of ST modified gravity for which the fits are the same or better than in Keplerian case. We then calculated orbits for the best fit parameters of ST gravity and compared them with observations. In that way, our results enable us to test ST theory at galactic scales.

In this paper we derived a particular theory among the class of ST theories of gravity. We tested this gravity theory by studying dynamics of S2 star around supermassive BH at GC. For 15 combinations of $m$ and $n$ parameters we obtain the values of $\xi$ and $\lambda$ for which S2 star orbits in ST gravity better fit astrometric observations than Keplerian orbit. We obtained much larger orbital precession for the best fit parameter values of the S2 star in ST gravity than the corresponding value predicted by GR. The precession of S2 star orbit has the positive direction, as in GR. Also, we discuss the Newtonian limit of this class of ST theories of gravity. We believe that the approach we proposed can be used to constrain the different modified gravity models from stellar orbits around GC (see also \cite{dela18a,dela18b,dial19}).

\FloatBarrier

 \bigskip

{\small\rm\baselineskip=10pt
 \baselineskip=10pt
 \qquad Vesna Borka Jovanovi\'{c}\par
 \qquad Atomic Physics Laboratory (040)\par
 \qquad Vin\v{c}a Institute of Nuclear Sciences, University of Belgrade\par
 \qquad P.O. Box 522\par
 \qquad 11001 Belgrade, Serbia\par
 \qquad {\tt vborka@vinca.rs}

 \bigskip \smallskip
 
 \qquad Predrag Jovanovi\'{c}\par
 \qquad Astronomical Observatory\par
 \qquad Volgina 7\par
 \qquad P.O. Box 74\par
 \qquad 11060 Belgrade, Serbia\par
 \qquad {\tt pjovanovic@aob.rs}
 
 \bigskip \smallskip

 \qquad Du\v{s}ko Borka\par
 \qquad Atomic Physics Laboratory (040)\par
 \qquad Vin\v{c}a Institute of Nuclear Sciences, University of Belgrade\par
 \qquad P.O. Box 522\par
 \qquad 11001 Belgrade, Serbia\par
 \qquad {\tt dusborka@vinca.rs}

 \bigskip \smallskip

 \qquad Salvatore Capozziello\par
 \qquad $^1$Dipartimento di Fisica "E. Pancini"\par
 \qquad Universit\`{a} di Napoli "Federico II"\par
 \qquad Compl. Univ. di Monte S. Angelo, Edificio G, Via Cinthia\par
 \qquad I-80126, Napoli, Italy\par
 \qquad $^2$Istituto Nazionale di Fisica Nucleare (INFN), Sez. di Napoli\par
 \qquad Compl. Univ. di Monte S. Angelo, Edificio G, Via Cinthia\par
 \qquad I-80126, Napoli, Italy\par
 \qquad $^3$Gran Sasso Science Institute, Viale F. Crispi, 7\par
 \qquad I-67100, L'Aquila, Italy\par
 \qquad {\tt capozzie@na.infn.it}
 
 \bigskip \smallskip
  
 \qquad Stefania Gravina\par
 \qquad $^1$Dipartimento di Fisica "E. Pancini"\par
 \qquad Universit\`{a} di Napoli "Federico II"\par
 \qquad Compl. Univ. di Monte S. Angelo, Edificio G, Via Cinthia\par
 \qquad I-80126, Napoli, Italy\par
 \qquad {\tt stefaniag.92@outlook.it}
 
  \bigskip \smallskip
  
 \qquad Anna D'Addio\par
 \qquad $^1$Dipartimento di Fisica "E. Pancini"\par
 \qquad Universit\`{a} di Napoli "Federico II"\par
 \qquad Compl. Univ. di Monte S. Angelo, Edificio G, Via Cinthia\par
 \qquad I-80126, Napoli, Italy\par
 \qquad {\tt anna.daddio2@studenti.unina.it} 
 }

 \end{document}